\documentclass[%
 reprint,
 amsmath,amssymb,
 aps,
]{revtex4-1}
\usepackage[utf8]{inputenc}
\usepackage{graphicx,subfigure}
\usepackage{xcolor}
\usepackage{siunitx}
\usepackage{balance}
\usepackage{amsmath}
\usepackage{amssymb}
\usepackage{breakcites}
\usepackage{ulem,lipsum}

\usepackage{tikz}
\usetikzlibrary{positioning,arrows}
\usetikzlibrary{decorations.pathmorphing}
\usetikzlibrary{decorations.markings}

\usepackage{bm}
\usepackage{textcomp}
\usepackage{float}

\newcommand\soutpars[1]{\let\helpcmd\sout\parhelp#1\par\relax\relax}
\long\def\parhelp#1\par#2\relax{%
  \helpcmd{#1}\ifx\relax#2\else\par\parhelp#2\relax\fi%
}
\begin{document}

\title{A helical-shape scintillating fiber trigger and tracker system for the DarkLight experiment and beyond}

\author{Yimin Wang}\email{yiminw@mit.edu}
\author{Ross Corliss}
\author{Richard G. Milner}
\affiliation{Laboratory for Nuclear Science, Massachusetts Institute of Technology, Cambridge, MA 02139, U.S.A}
\author{Christoph Tschal\"{a}r}
\affiliation{
  MIT Bates Research and Engineering Center, Middleton, MA 01949, U.S.A}
\author{Jan C. Bernauer}
\affiliation{Stony Brook University, Stony Brook, NY 11794, U.S.A}
\affiliation{RIKEN BNL Research Center, Brookhaven National Laboratory, Upton, NY 11973, U.S.A}

\date{\today}

\begin{abstract}
    The search for new physics beyond the Standard Model has interesting possibilities at low energies. 
    For example, the recent 6.8$\sigma$ anomaly reported in the invariant mass of $e^+e^-$ pairs from $^8\text{Be}$ nuclear transitions and the discrepancy between predicted and measured values of muon g-2 give strong motivations for a protophobic fifth-force model. 
    At low energies, the electromagnetic interaction is well understood and produces straightforward final states, making it an excellent probe of such models.
    However, to achieve the required precision, an experiment must address the substantially higher rate of electromagnetic backgrounds.
    In this paper, we present the results of simulation studies of a trigger system, motivated by the DarkLight experiment, using helical-shape scintillating fibers in a solenoidal magnetic field to veto electron-proton elastic scattering and the associated radiative processes. 
    We also assess the performance of a tracking detector for lepton final states using scintillating fibers in the same setup.
\end{abstract}

\maketitle

\section{Introduction}
The search for new physics beyond the Standard Model is a major research thrust for physicists worldwide. The energy frontier is pursued at the Large Hadron Collider and the cosmic frontier is explored in searches for dark matter with underground experiments. 
High intensity experiments at low energy offer a powerful and complementary approach.  These experiments can directly investigate recent low energy anomalies, e.g. the 6.8$\sigma$ anomaly which has been reported in $^8\text{Be}$ nuclear transitions. The invariant mass of the $e^+e^-$ pairs from $^8\text{Be}$ transitions shows an unexpected excess at around \SI{17}{MeV}\cite{Krasznahorkay2016}. A protophobic fifth-force model with a $\sim$\SI{17}{MeV} vector gauge boson has been proposed to explain this anomaly\cite{Feng:2016ysn}. This model can also address the 3.6$\sigma$ discrepancy between the predicted and measured muon's anomalous magnetic moment\cite{Bennett2006}. Furthermore, this vector gauge boson can also be a dark matter candidate as the interaction mediator between the dark sector and the visible sector.

The simplest fifth-force model is the so-called minimally kinetically mixed model\cite{Holdom1986}, where a dark photon $A'$ couples to the standard model through the vector portal.  It inherits electromagnetic couplings through this mechanism, with a reduced strength $\alpha'=\epsilon^2\alpha_\text{EM}$, where $\alpha_{\text{EM}} \approx1/137$. 
This and other fifth force models are generally described by the mediator mass $m_{A'}$ and one or more couplings $\alpha'$.

The simple model has been ruled out to $\sim 2\sigma$ for the muon and $^8$Be anomalies.
However, more general models with flavor-dependent couplings are consistent with all current exclusion limits \cite{Feng:2016ysn} and could explain these anomalies.  Such a new force could be seen in low-energy electroproduction via decay to $e^+e^-$ final states.

\section{DarkLight experiment}

The DarkLight experiment\cite{Balewski2014} proposes to search for new physics beyond the Standard Model via decays to $e^+e^-$ final states in the mass region from \SI{10}{MeV/c^2} to \SI{100}{MeV/c^2}. The experiment passes an electron beam through a windowless, cylindrical hydrogen gas chamber of length of \SI{60}{cm}. The gas target has low density to allow the recoil proton to propagate to a detector at the chamber wall to be measured. The extended length of the target along the beam line gives a target thickness of \SI{e19}{cm^{-2}}. We focus at low beam energy of \SI{100}{MeV} and measure the complete final-state in $e+p\rightarrow e+p+e^++e^-$. The detector is located within a \SI{0.5}{T} solenoidal magnet for momentum measurements. 

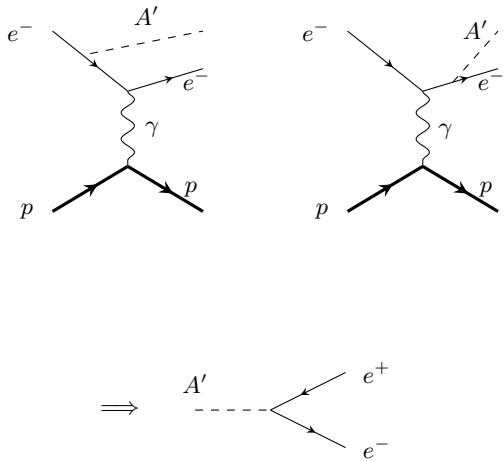
\begin{figure}[h!]
\centering
 \tikzset{     
    photon/.style={decorate, decoration={snake}},    
    electron/.style={draw=black, postaction={decorate},
    decoration={markings,mark=at position .6 with {\arrow[black]{stealth}}}},
    aprime/.style={dashed},
    proton/.style={very thick, postaction={decorate}, decoration={markings,mark=at position .6 with {\arrow[black]{stealth}}}}
     }

\begin{tikzpicture}[]
    \draw[electron] (-1,1.2) node[label=left:$e^-$] {} --( 0,0.4);
    \draw[electron] ( 0,0.4) -- node[label=right:$e^-$] {} (1,0.7);
    \draw[aprime] (-0.5,0.9) -- node[label=above:$A'$] {} (1,1.2);
    \draw[proton] (-1,-1.2) node[label=left:$p$] {} -- ( 0,-0.6);
    \draw[proton] (0,-0.6)-- node[label=right:$p$] {} (1,-1.2);
    \draw[photon] (0,0.4) -- node[label=right:$\gamma$] {} (0,-0.6);
\end{tikzpicture}\hskip1cm
\begin{tikzpicture}[]
    \draw[electron] (-1,1.2) node[label=left:$e^-$] {} --( 0,0.4);
    \draw[electron] ( 0,0.4) -- node[label=right:$e^-$] {} (1,0.7);
    \draw[aprime] (0.4,0.52) -- node[label=above:$A'$] {} (1,1.2);
    \draw[proton] (-1,-1.2) node[label=left:$p$] {} -- ( 0,-0.6);
    \draw[proton] (0,-0.6)-- node[label=right:$p$] {} (1,-1.2);
    \draw[photon] (0,0.4) -- (0,-0.6) node[label={[label distance=0.15cm]70:$\gamma$}] {} ;

\end{tikzpicture}\hskip1cm\begin{tikzpicture}[]
    \draw (-2,0) node{$\Longrightarrow$};
    \draw[aprime] (-1,0) node[label=above:$A'$] {} -- (0,0);
    \draw[electron] (1,0.5) node[label=right:$e^+$] {} --( 0,0.) ;
    \draw[electron] ( 0,0.) -- (1,-0.5) node[label=right:$e^-$] {};
    \useasboundingbox (-1,-1.2) rectangle(2,2.4);
\end{tikzpicture}
\caption{Feynman diagrams for the signal processes: The $A^\prime$ is produced off the incoming or outgoing lepton and then decays into an $e^+e^-$ pair.}
\label{fig:feyn}
\end{figure}

In these collisions, a hypothetical force mediator, $A'$, is produced via the ``$A'$-strahlung" processes. The $A'$ then decays into an $e^+e^-$ pair, which is recorded in the detectors. Figure~\ref{fig:feyn} shows these production processes for signal. There are also irreducible QED background processes which generate an $e^+e^-$ pair via an intermediate virtual photon. At the design luminosity, however, the dominant background processes in the DarkLight experiment are elastic and Møller scattering and their associated radiative processes on which this study focuses. Although these processes have different final states from the signal processes, their overwhelming rate at our luminosity prohibits directly triggering on these background processes.

This report summarizes the study of a trigger system based on helical-shape scintillating fibers. 
Using a Geant4\cite{Geant4} simulation of the physics and detector design, we  show that this trigger system can filter out most of the electron-proton elastic scattering. 
We also explore the possible use of scintillating fibers as the basis for the lepton tracking detectors. 
Though the specific design in this study is motivated by the DarkLight experiment, the general principles can be easily extended for other experiments using intense electron beams with similar energy.

\section{Fiber trigger system}
The signal event $e+p\rightarrow e+p+e^++e^-$ contains one recoil proton and three leptons which the system will trigger on. However, the electron-proton elastic scattering and its associated radiative processes produce overwhelming count-rates in the detector. Therefore, we cannot trigger on the number of leptons alone. In order to read out the detector in this environment, the trigger must be made sufficiently blind to elastic scattering events. The proposed trigger system contains two pieces of logic: one identifies the background electron-proton elastic scattering events by virtue of their characteristic scattering geometry and ignores these events; the other one requires three simultaneous leptons accompanying one identified proton to reflect the prompt decay of an $A'$ into $e^+e^-$ pair.

\subsection{Background kinematics and trigger system geometry}

\begin{figure}[h!]
\centering
\includegraphics[width=0.9\linewidth]{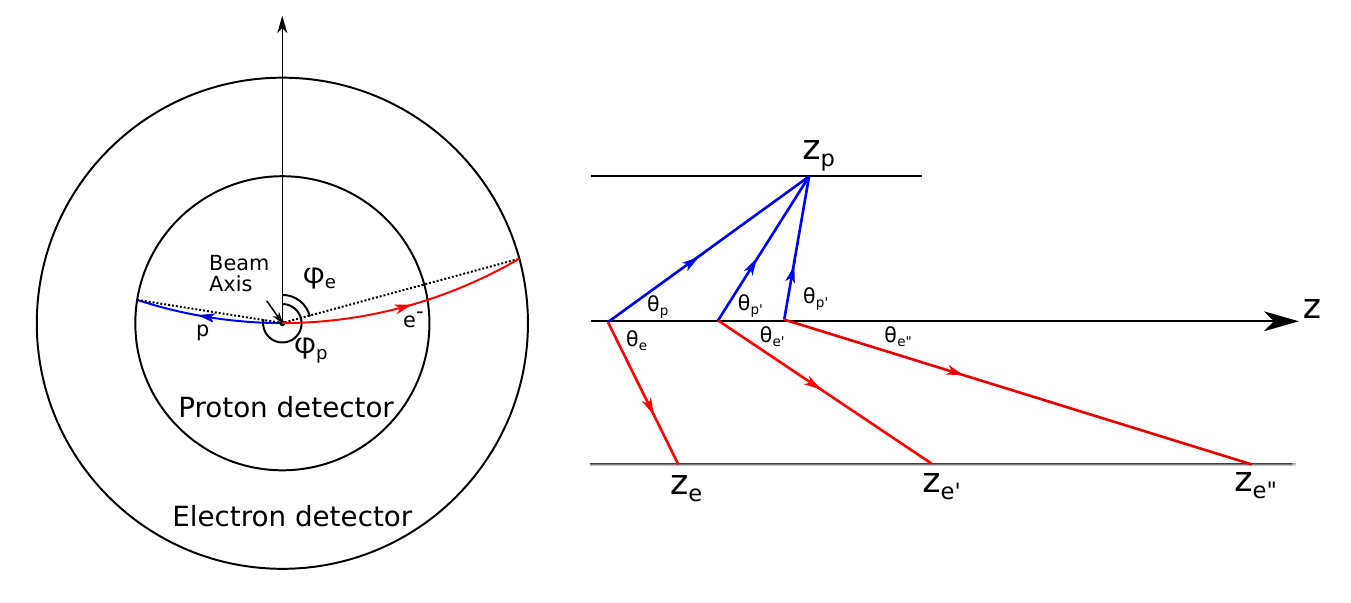}
  \caption{Scattering geometries: Transverse (Left) and longitudinal (Right) scattering geometries show the definition of relevant quantities, which are described in detail in the text. Because we use $\Delta \phi$ and $\Delta z$, the zero point for $\phi$ and $z$ are irrelevant. Notice that, in this design, $R_p \lessapprox R_0 \lessapprox R_e$, where $R_0$ is the radius of the target chamber wall. The difference in radii is exaggerated for better illustration.}
  \label{fig:scatterGeo}
\end{figure}

The cylindrical target chamber of radius $R_0=$\SI{75}{mm} is centered on the beam axis along the $z$ direction. The trigger detector consists of a sensitive layer for protons just inside the wall of the chamber and a sensitive layer for electrons just outside it. In elastic events, the scattered proton and electron reach the trigger detector located at a fixed radius $R_0$ at different azimuth angle $\phi$ and position $z$, namely ($\phi_p$, $z_p$) and ($\phi_e$, $z_e$), as shown in Figure~\ref{fig:scatterGeo}.  Because the target has extended length along the beam line, the absolute values of these two quantities depend not only on the kinematics but also on the location of the primary vertex along the beam path. 

However, we can define the following two quantities whose values depend only on the kinematics but not on the location of the primary vertex: 
\begin{equation*}
\centering
\begin{aligned}
\Delta z & \equiv z_e - z_p \\
\Delta \phi & \equiv \phi_e-\phi_p  \pmod{2\pi}
\end{aligned}
\label{eqn:deltaphiz}
\end{equation*}

If an elastically scattered electron and proton have high enough transverse momenta to reach the tracking detector, their transverse path lengths within the target chamber can be approximated by the radius $R_0$ of the chamber.
This approximation, combined with numerical calculations of these two quantities, reveals a nearly linear correlation between $\Delta z$ and $\Delta \phi$ in the range of \SI{-50}{mm}$\leq \Delta z \leq$ \SI{150}{mm} covered by the detector for our beam energy and chamber radius as shown in Figure~\ref{fig:ZPhiScatter}. 

In the case of a recoil proton hitting the trigger detector at position ($\phi_{p0}$,z$_{p0}$), according to this linear relation, the electron hit lies on a helix whose position can be calculated from the proton hit position ($\phi_{p0}$,$z_{p0}$).  The position of the electron hit along this helix depends on the position of the scattering vertex. Similarly, for any scattered electron which hits this helix, this linear relation indicates that the recoil proton also strikes a corresponding helix that passes through ($\phi_{p0}$,z$_{p0}$). Therefore, elastic events will fall on known pairings of helices regardless of the position of the scattering vertex along the beam path.

\begin{figure}[h!]
\includegraphics[width=0.9\linewidth]{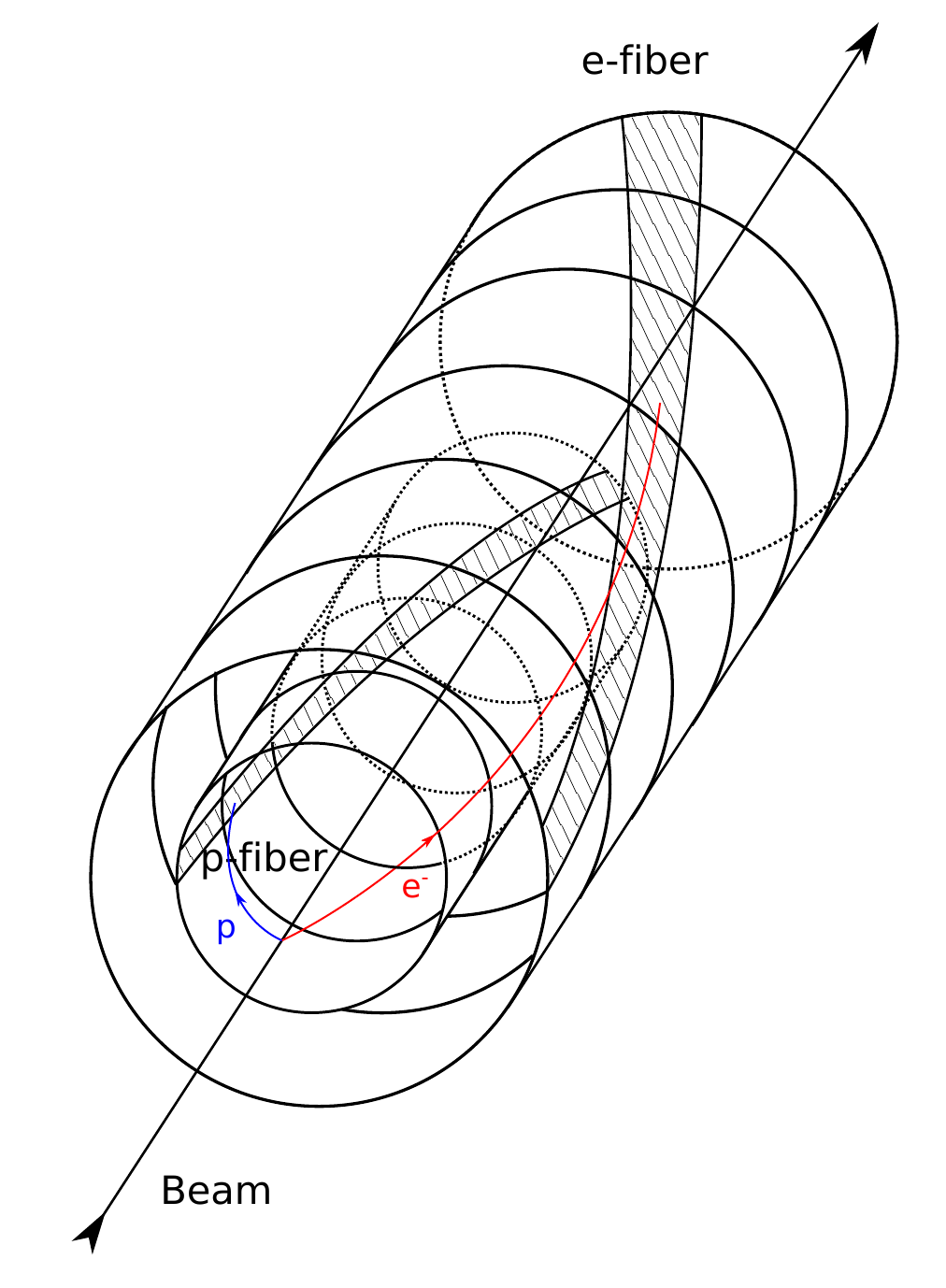}
  \caption{Fiber geometries: One pair of elastically scattered proton and electron hit the paired electron fiber and proton fiber. In our design, $R_p \approx R_e \approx$ \SI{75}{mm}. The radius difference in the drawing is exaggerated for better illustration.}
  \label{fig:fiberLayout}
\end{figure}

The linear relation at a small fixed radius characterizes most elastic events and can thus be used to suppress triggering on the majority of the elastic electron-proton scattering background. We propose to use scintillating fibers in the shape of helical strips as trigger detectors to cover the cylindrical surface of radius $R_0=$\SI{75}{mm} as shown in Figure~\ref{fig:fiberLayout}. Each fiber for detecting recoil protons has a corresponding fiber for detecting scattered electrons paired to it. If one proton fiber fires in coincidence with its elastic partner electron fiber, this is ignored. If one proton fiber fires, but NOT the partner electron fiber, we keep this event at this stage, and judge with more information at the next stage. This completes the first part of the trigger logic.

The second part of the trigger logic requires that at least three electron hits remain after the elastic pairs are removed. Otherwise, this event is either a false positive from the previous step or a signal event outside of the trigger acceptance.

\subsection{Event simulation}
The simulation uses Geant4 to construct the full geometry of the experiment layout. We choose proton fibers to cover a length of \SI{23}{cm} and electron fibers to cover a length of \SI{46}{cm} along the beam direction. These dimensions guarantee the full coverage of the momentum phase space for our experiment. The background events are simulated elastic scattering with radiative corrections, assuming a proton target and a \SI{100}{MeV} beam. We generate these events with proton recoil angles covering the trigger detector acceptance with an additional 3\textdegree\ margin on both sides. The primary vertex locations are uniformly distributed along the path the beam takes through the target. These generated events include all reconstructible events but do not guarantee the generated proton will hit a proton fiber when the primary vertex is far downstream in the target chamber. 

The signal events are generated with MadGraph\cite{Alwall:2014hca} for $A'$ masses from \SI{10}{MeV} to \SI{90}{MeV}\cite{Kahn2012}. We generate a primary vertex position for each event the same way as for the background events. We then select and focus on signal events in which all final state particles hit the fibers and thus give enough information for event reconstruction. These events provide a proper basis to evaluate the signal efficiency of our trigger system. We include $10^6$ background events and $3\times10^4$ signal events to limit the effects from statistical fluctuations. 

\begin{figure}[h!]
\centering
\includegraphics[width=0.9\linewidth]{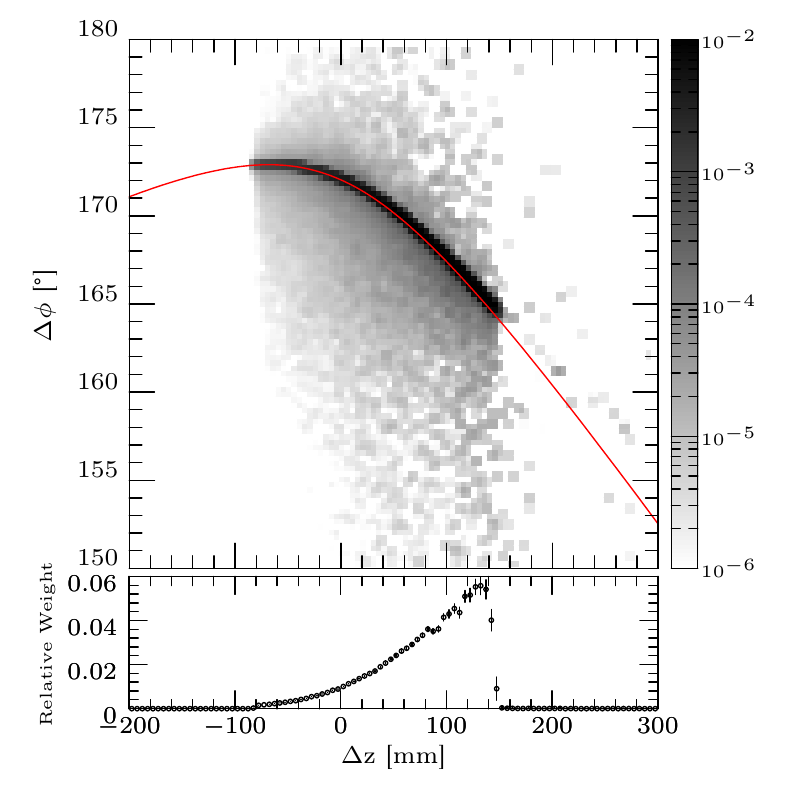}
  \caption{Upper panel: $\Delta\phi$ and $\Delta z$ scatter plot compared with the analytical calculation along the red line. Lower panel: histogram of relative weight of cross-section of $\Delta z$. The abrupt cut-offs at $\Delta z=$\SI{-90}{mm} and \SI{150}{mm} are from the limited angle simulated, and corresponds to the acceptance of the electron layer.}
\label{fig:ZPhiScatter}
\end{figure}

The $\Delta z - \Delta\phi$ scatter plot in Figure~\ref{fig:ZPhiScatter} shows that the simulation agrees with theoretical calculations for the electron-proton elastic scattering background. It also verifies that most elastic background events fall into the linear region. 

\begin{figure}[h!]
\centering
  \includegraphics[width=0.9\linewidth]{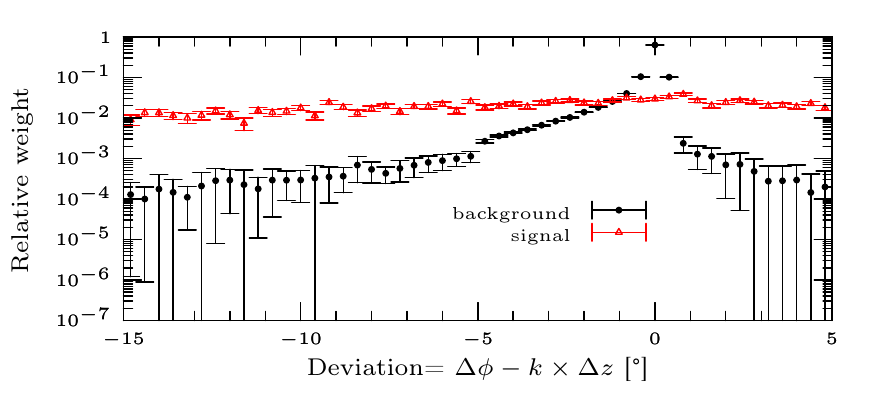}
  \caption{Deviation histogram: The histogram shows the relative weight of cross-section of deviations. The deviation is defined as the difference between simulated $\Delta \phi$ and calculated $\Delta \phi$ using simulated $\Delta z$ and the linear fit from the $\Delta\phi$ and $\Delta z$ scatter plot. The black dots are from elastic background, and red dots from signal of $A'=$\SI{20}{MeV}. The abrupt drops on both ends of the deviation spectra for elastic background are from the kinematic cut on the detectable electron scattering angle in background event generation.}
  \label{fig:dev}
\end{figure}

We parameterize the linear region of this scatter plot via the equation $\Delta\phi=k\times\Delta z+\phi_0$. We then perform a linear fit in the region of \SI{20}{mm} $\leq \Delta z \leq$ \SI{140}{mm} to get parameters $k$ and $\phi_0$. We define the deviation from this line for each event as the difference between the simulated $\Delta \phi_{sim}$ value and the calculated $\Delta\phi_{cal}=k\times\Delta z_{sim}+\phi_0$ value using parameters from the linear fit. The histogram of this deviation value for both the background events and the signal events is shown in Figure~\ref{fig:dev}. The background events cluster around zero deviation, while the signal events are distributed nearly uniformly (we plot only the smallest deviation among all three leptons in each signal event).

\begin{figure}[h!]
\centering
\includegraphics[width=0.9\linewidth]{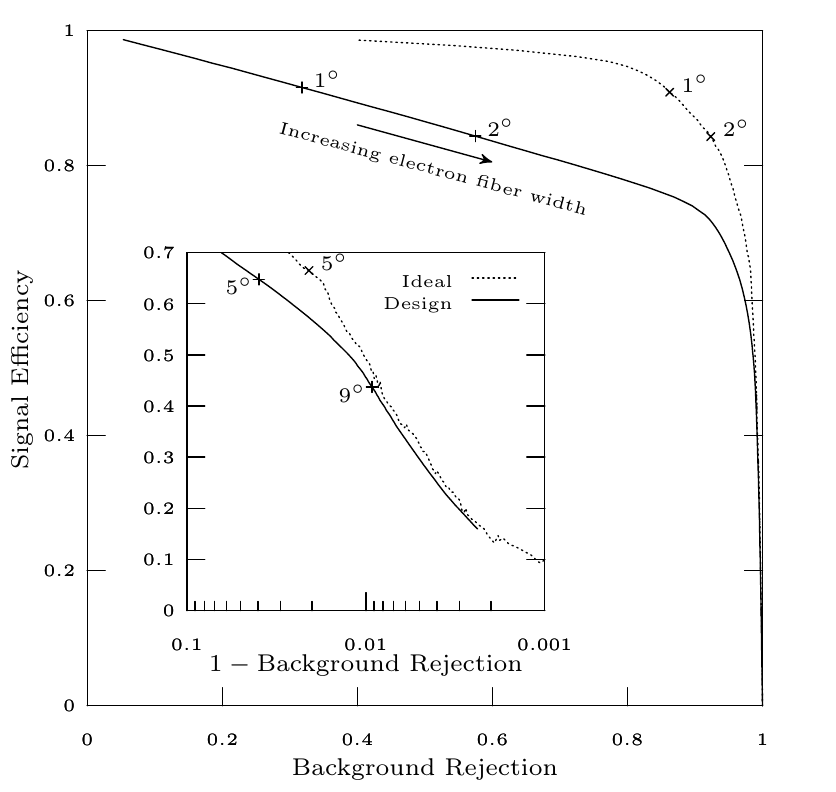}
  \caption{Signal efficiency vs background rejection as the width of the electron fiber is varied: The dotted line is for an infinitely narrow proton fiber assumption and the solid line for the 3.6\textdegree\ wide proton fiber. The inset shows the behavior near 100\% background rejection, shown as 1-BR. Several points are marked on these two lines to show the corresponding electron fiber width. Background rejection is optimized by varying the electron fiber offset for different widths of the paired electron fiber. Both numbers drop as we increase the electron fiber width which excludes more phase space.}
\label{fig:ROC}
\end{figure}

The idealized scenario uses an infinitely narrow proton fiber, but a real detector must have a finite width. This introduces a finite resolution in the $\Delta \phi$ and $\Delta z$ measurement. Our design uses 100 proton fibers, each covering 3.6\textdegree\ of azimuth angle (corresponding to a fiber width of \SI{4.7}{mm}). Because the deviation value from the background events is asymmetric, we optimize the background rejection for different electron fiber widths by adjusting $\phi_0$ offsets between each pair of proton fiber and electron fiber. The background rejection and signal efficiency for different electron fiber widths are shown in Figure~\ref{fig:ROC}. We pick electron fiber width of 7.2\textdegree\ (\SI{9.4}{mm}) which gives approximately 98\% background rejection for our design. Two layers, each with 50 channels of electron fibers, are used for one-to-one pairing to proton fibers. This design should be able to filter out most of the background events, leaving only the rare case of background from multiple scattering and secondary particle production.

\subsection{Full system performance}
The previous section verifies the analytical calculations and summarizes the reasoning behind our choices for the fiber parameters. This section further evaluates this trigger system with more detailed simulation. There are two important factors added into the full simulation:

\begin{enumerate}
  \item Physical fiber geometry:\\
    The fibers we studied have a square cross section, and carry cladding, which forms non-sensitive regions. We assume \SI{0.5}{mm} thick fibers, with a typical 5\% thickness of cladding on all surfaces. The non-sensitive regions of electron fibers will result in elastic electrons not detected and thus leads to possible false positives. 

  \item Event pile-up:\\ 
     We assume a reasonably achievable readout rate of \SI{10}{MHz}. This results in multiple elastic events in every time-frame for the beam from a typical linac. We use Poisson statistics to simulate the pile-up to study the effects on the trigger system. We assume the beam current to be \SI{5}{mA} with \SI{75}{MHz} bunches and a coincidence window of \SI{100}{ns}. There are on average 110 background events in each time frame. For the signal efficiency study, an additional $A'$ event is combined with these pile-up background events in each time-frame to study.
\end{enumerate}

These two factors are predictable and readily simulated aspects of a physical implementation.  While other details such as light attenuation and readout noise are also important to consider, their extent is more dependent on the specifics of the implementation, and so they are not treated in detail here.

In the following, we refer to one time frame as one event. The trigger procedure can be broken down into the following steps:

\begin{figure}[h!]
\centering
\includegraphics[width=0.9\linewidth]{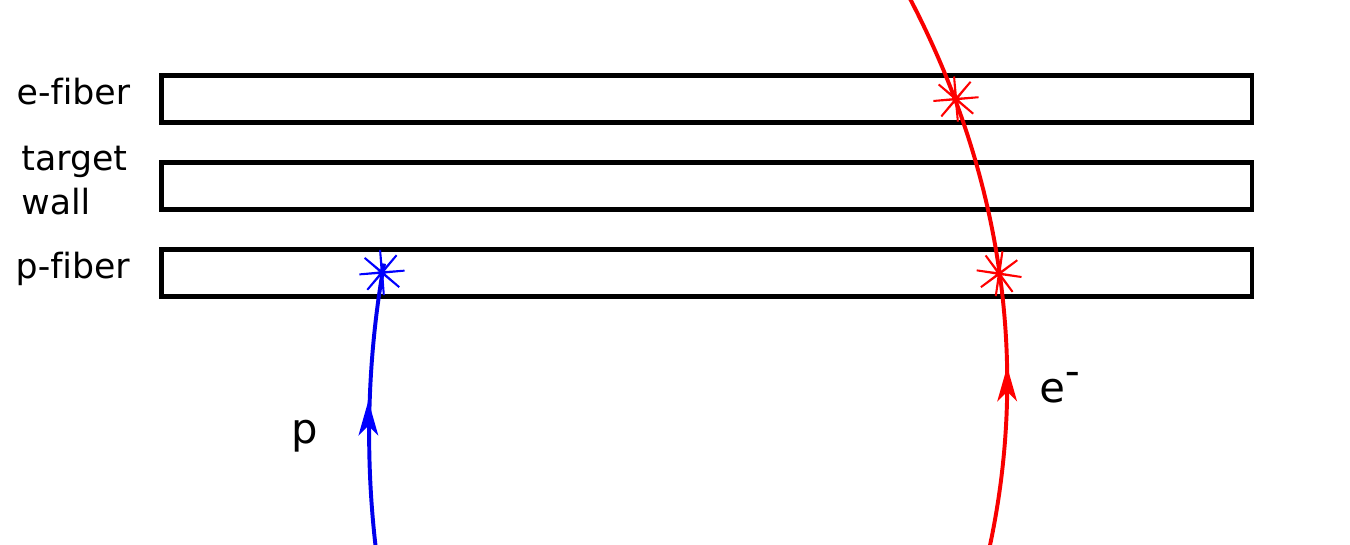}
  \caption{Proton and electron hits in fibers: Protons are stopped by proton fibers. Electrons deposit energy in both proton fibers and electron fibers with the target wall of \SI{2}{mm} aluminum in between.}
\label{fig:protonID}
\end{figure}

\begin{figure}[h!]
\centering
\includegraphics[width=0.9\linewidth]{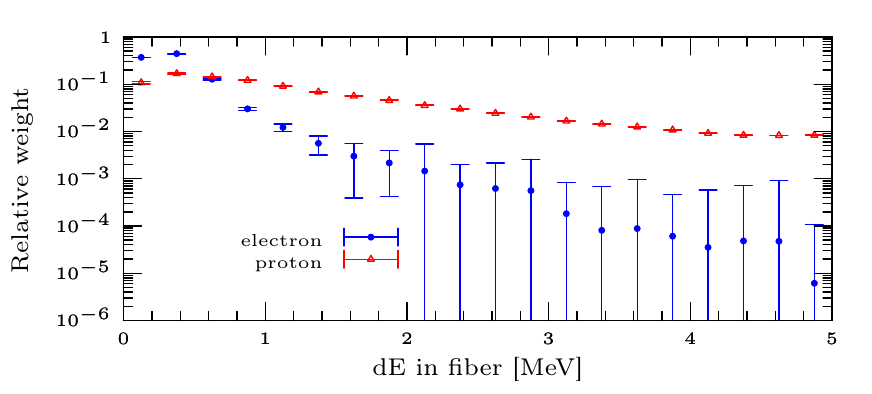}
  \caption{Histogram of the relative weight of proton(red) and electron(blue) energy deposition in \SI{0.5}{mm} thick plastic fibers. The proton dE is very similar to the recoil energy spectrum, since the protons generally stop in the fiber. The low energy region (dE\textless \SI{0.5}{MeV}) is dominated by the typical energy deposition from electrons.}
\label{fig:energyDeposition}
\end{figure}

\begin{itemize}
    \item Particle identification\\
    Due to their high dE/dx and low momentum, scattered protons deposit all their energy in the \SI{0.5}{mm} thick plastic fibers inside the chamber wall, i.e. the proton fibers. However, most scattered electrons propagate as MIPs, depositing an energy almost independent of their momenta in the fibers. These scattered electrons generally penetrate the target chamber wall and deposit energy in both proton fibers and electron fibers, leaving a very different signature compared to protons, as shown in Figure~\ref{fig:protonID}. Therefore, the proton identification utilizes information from both fibers: the energy deposition in proton fibers and the anti-coincidence with energy deposited in nearby electron fibers. 
    
    Based on the simulated energy deposition spectrum for elastic scattering, as shown in Figure~\ref{fig:energyDeposition}, a proton-like hit is defined by an energy deposition \textgreater $E_p$=\SI{0.5}{MeV} in a proton fiber and an energy deposition \textless \SI{0.1}{MeV} (set above zero to allow for low-energy noise) in the electron fiber geometrically behind it. An lepton-like hit is defined by by an energy deposition \textgreater \SI{0.1}{MeV} in both a proton fiber and the electron fiber geometrically behind it. 
    
    Setting a higher $E_p$ in proton fibers can help eliminate events with low energy electrons that hit one proton fiber and are either stopped by the target chamber wall or go through the cladding of electron fibers and thus deposit no energy in the electron fibers. This specific value, $E_p$=\SI{0.5}{MeV}, might change depending on the noise and signal attenuation in the real experiment, but can be optimized by scanning threshold values.  As $E_p$ decreases, protons will begin to trigger the detector, followed by a much sharper rate increase when electrons become visible. 
    \item Background rejection\\
    After the particle identification, all signals from paired proton and electron fibers hit by elastic events are neglected. This leaves us with hits on the remaining fibers from non-elastic events or elastic events at the non-linear tail of low $\Delta z$. It also produces false negatives for non-elastic events in which there is a low-energy scattered electron with close to zero deviation in $\Delta \phi - \Delta z$ phase space. 
    \item Signal trigger\\
    After hits from elastic backgrounds are rejected, the trigger system triggers when there is at least one remaining proton hit and at least three lepton hits. The surviving events from background are rare false positives from the non-linear elastic tail coinciding with multiple low energy scattered electrons.
\end{itemize}

\begin{figure}[h!]
\centering
\includegraphics[width=0.9\linewidth]{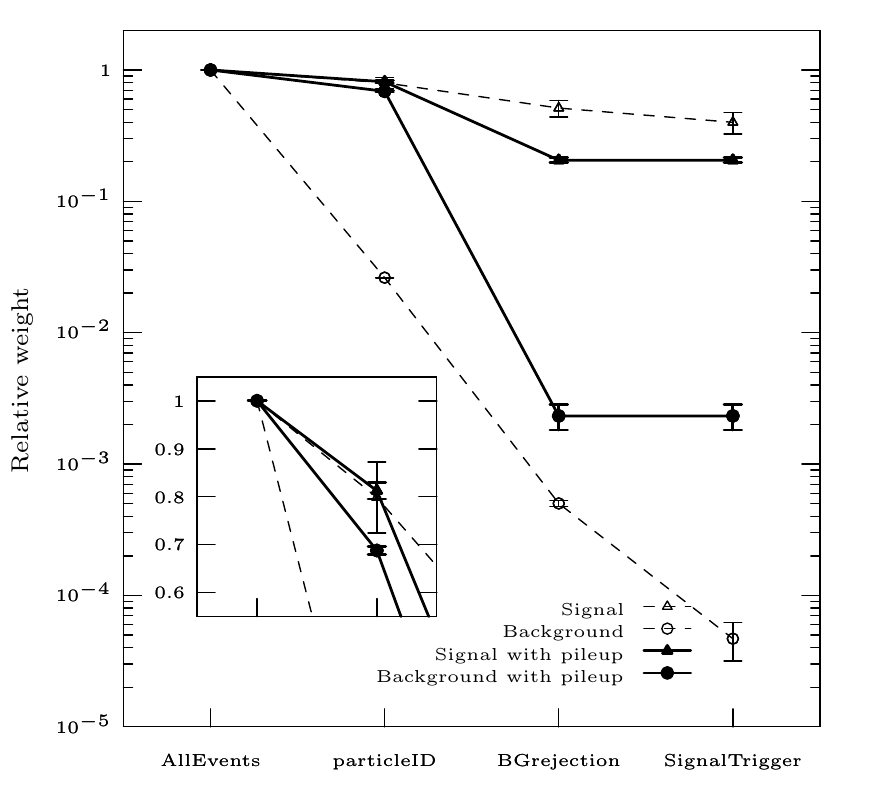}
\caption{Simulated trigger results: Relative weight of background events and signal events of $A'$ =\SI{20}{MeV}, with and without simulated pile-up, on each stage of trigger logic. The pile-up level is based on a beam current of \SI{5}{mA}, and the assumption of a \SI{100}{ns} coincidence window. The results without pile-up are drawn with dashed lines and hollow marks and the results with pile-up are drawn with solid lines and marks. Pile-up introduces multiple protons and multiple electrons in each time-frame, limiting the performance of proton ID and the triple-lepton trigger stage.}
\label{fig:cut}
\end{figure}

Trigger performance without event pile-up is shown by dashed lines in Figure~\ref{fig:cut}. The protonID step rejects events that have no proton-like hits. Background events rejected at this step are those with primary vertex far enough downstream that the proton is outside the acceptance. For signal events, this is already in consideration when we select signal events, the remaining small loss at this step is from proton misidentification. Removing hits in the electron fibers paired with struck proton fibers cuts the elastic background down by two orders of magnitude, as expected by the design. This has little impact on the signal events except for random coincidence. The requirement for remaining proton and lepton hits cuts the background events by another large factor while still keeping most signal events. 

Trigger performance with event pile-up is shown by solid lines in Figure~\ref{fig:cut}. Multiple elastic background events in each frame almost guarantee a proton hit and also introduce more electron hits without paired proton hits from secondary processes. Increased numbers of electron hits in each time-frame compared to single event increase the accidental pairing of proton and electron fibers, causing larger signal loss at this step. It also almost guarantees at least three unpaired electron hits in each time-frame. Therefore, the trigger requirement of three lepton hits loses all its effect. 

We assume the beam current to be \SI{5}{mA} and a coincidence window of \SI{100}{ns} for our pile-up study in Figure~\ref{fig:cut}. At this readout speed, the trigger system reduce the trigger rate to around \SI{23}{kHz} while achieve a signal efficiency $\sim$25\%. 
We also explored a scenario with half the pile-up, corresponding to a \SI{50}{ns} coincidence window, or a reduced luminosity.  In this scenario, the resulting signal efficiency is 35\% and the false triggering ratio is slightly decreased compared to the \SI{100}{ns} baseline.

Because the kinematic quantities $\Delta z$ and $\Delta \phi$ used in the trigger system have no dependence on $A'$ mass in signal events, the signal efficiency for this trigger system is very similar for signal events across the whole $A'$ mass region we study from \SI{10}{MeV} to \SI{90}{MeV}. 

\subsection{Summary}
Our study confirms that even with several confounding factors, such a trigger system is capable of dramatically reducing the trigger rate compared to simply triggering on the presence of three leptons in a time-frame. At a beam current of \SI{5}{mA}, and assuming a \SI{100}{ns} coincidence window, this fast trigger system can reduce the trigger rate to around \SI{23}{kHz}. However, at this readout speed, pile-up reduces the signal efficiency to $\sim$25\%. 
Reduced luminosity or a smaller coincidence window would decrease pile-up, yielding nonlinear improvement in the signal efficiency and lower false triggering ratio, which the study of the \SI{50}{ns} coincidence window scenario confirms.

\section{Scintillating Fiber Lepton Tracker}
\subsection{Concept and simulation}
\begin{figure}[h!]
\centering
\includegraphics[width=0.9\linewidth]{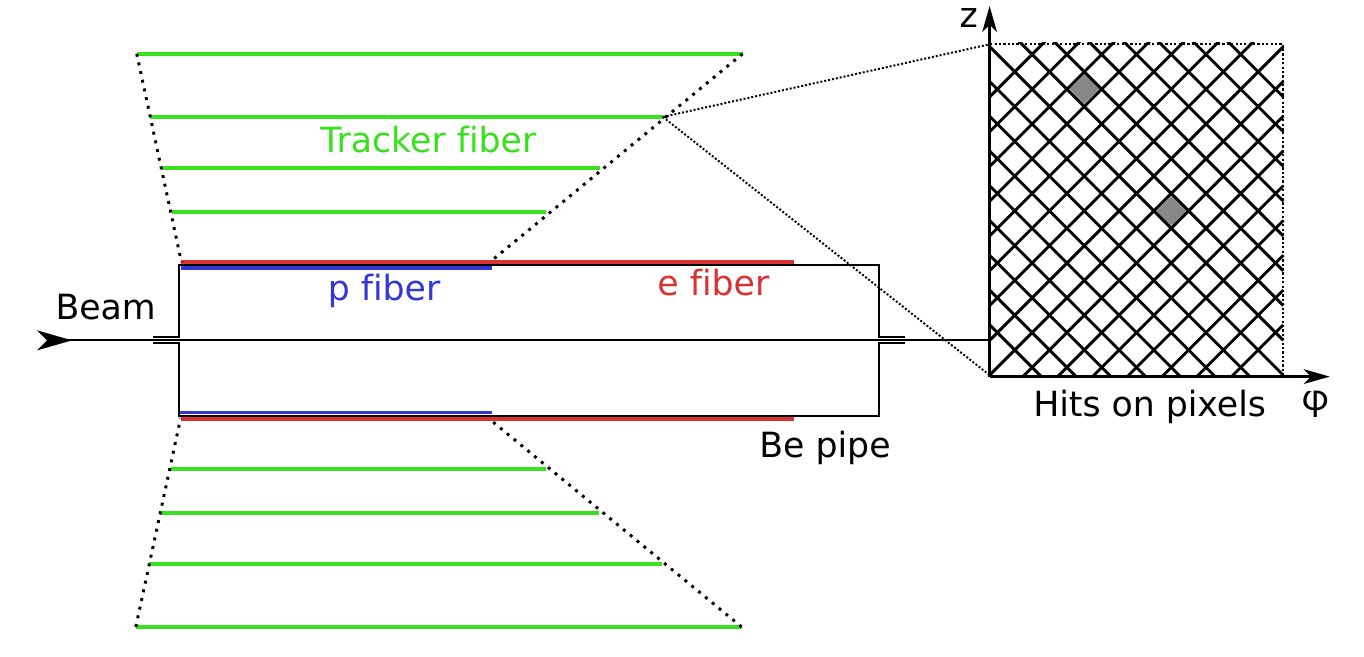}
\caption{Tracker position relative to the gas target chamber and the helical fiber trigger system: Green lines represent cylindrical lepton detector layers.  Each cylindrical shell contains two layers of fibers similar to Figure~\ref{fig:fiberLayout} with $\pm 45 $\textdegree\ relative to cylindrical axis, at right angle with each other. The drawing on the right shows the layout of fibers on the cylindrical surface of each tracker shell. Coincidence of hits in the perpendicular fibers in each layer provides position information.}
\label{fig:tracker}
\end{figure}

In addition to the trigger, we also study helical fibers as a possible cost-effective tracker. As shown in Figure~\ref{fig:tracker}, the tracking system we consider consists of four cylindrical shells surrounding the target chamber. They have increasing length in the beam direction at larger radius to cover the angular range for leptons. Each shell has two layers of fibers with a structure similar to Figure~\ref{fig:fiberLayout} but at right angles to each other in the cylindrical plane, forming square overlaps. A lepton lights up one pair of crossing fibers at each radius and gives position information to reconstruct the track. 

We design each fiber to be $45^\circ$ from the beam axis in cylindrical coordinates. There is currently no native geometry in Geant4\footnote{Geant4 version 10.3} that that can adequately describe this geometry, so, for the tracker system, each layer of fibers is coded as a cylindrical shell in Geant4 simulation. The fiber geometry and the inactive cladding region between fibers are later simulated separately based on the Geant4 output.

In our design, four shells have radii of \SI{100}{mm}, \SI{150}{mm}, \SI{200}{mm} and \SI{250}{mm}. The lengths along the beam direction increase linearly from \SI{230}{mm} for the innermost one to \SI{624}{mm} for the outermost one. Each layer has 1000 fibers with a thickness of \SI{0.5}{mm} and widths increasing from \SI{0.9}{mm} to \SI{2.2}{mm}. The dominant uncertainty in measurements at larger radius comes from multiple scattering of outgoing leptons. Therefore, we choose to have the same number of fibers for all shells and allow lower position resolution using wider fibers at larger radius.

\subsection{Performance}
We evaluate the performance of this design by studying the momentum resolution using single simulated leptons. Then we use this momentum resolution to infer the invariant mass resolution of $A'$ from lepton pairs. Finally, we plot the reach of our experiment using this detector design based on the $A'$ mass resolution.

We simulate electrons with longitudinal and transverse momenta both from \SI{5}{MeV/c} to \SI{50}{MeV/c} in \SI{5}{MeV/c} steps. These electrons start from the same vertex and propagate through each layer of fibers, leaving energy in a certain set of fibers. We extract their hit positions as the centers of those square regions formed by the coincidence of two layers of overlapping fibers. Finally, we apply the Karim{\"a}ki method\cite{Karimaki1991} for circular trajectory fitting on these hit positions to reconstruct the momentum. 

Momentum resolution is limited by intrinsic position resolution from the fiber width, and by the fact that the Karim{\"a}ki method does not provide corrections for energy loss and multiple scattering.

\begin{figure}[h!]
\centering
\includegraphics[width=0.9\linewidth]{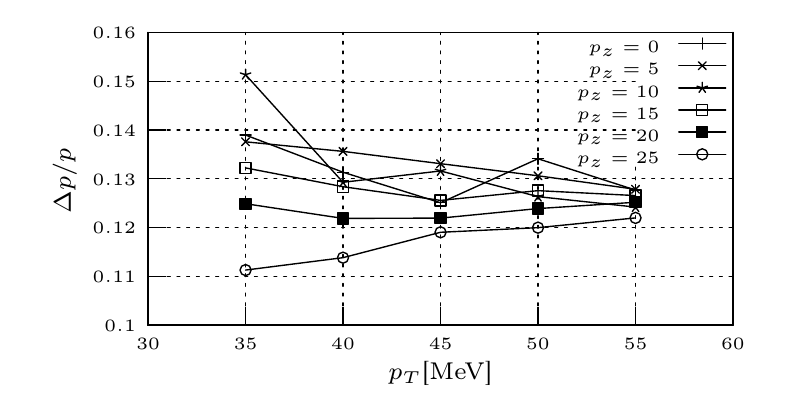}
\caption{Momentum resolution from the fiber tracker with the Karim{\"a}ki method over simulated momentum phase space: $\Delta p$ is defined as the magnitude of the vector difference between $p_{sim}$ and $p_{recon}$. The simulation is based on a \SI{2}{mm} thick aluminum chamber wall and a total of \SI{1.5}{mm} thick veto fibers.}
\label{fig:resolution}
\end{figure}

The simulation gives a rough map of resolutions in the momentum phase space we are interested in. Using this resolution, we then smear the momentum of leptons in the simulated signal events and estimate the invariant mass resolution.  
With the current target chamber design, which has a \SI{2}{mm} thick aluminum wall and a total of \SI{2}{mm} thick fiber for background vetoing, the momentum resolution is about 13\% across all momenta simulated as shown in Figure~\ref{fig:resolution}, leading to an invariant mass resolution around 20\% for $A'$=\SI{20}{MeV}. We also simulated a very aggressive design with \SI{10}{\mu m} thick Havar target chamber wall and \SI{100}{\mu m} thick scintillating fibers which yields a momentum resolution of about 7\% and an invariant mass resolution around 15\% for $A'$=\SI{20}{MeV}. These resolutions might be further improved with a more sophisticated tracking algorithm.  Optimizing the tracking, especially in light of the large amounts of pile-up, is beyond the scope of this study.


\begin{figure}[h!]
\centering
\includegraphics[width=0.9\linewidth]{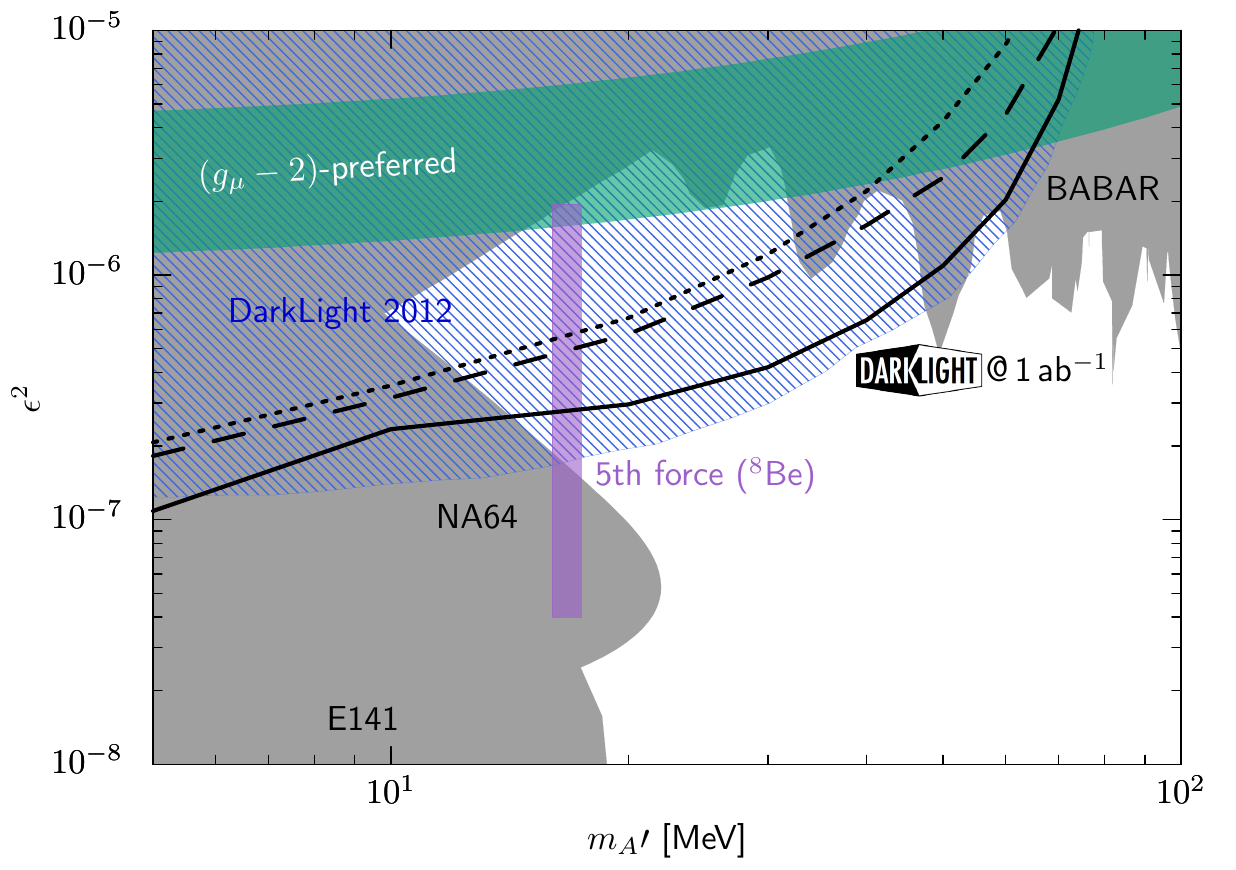}
\caption{Projected reach plot: 2$\sigma$ reach for an integrated luminosity of \SI{1}{ab^{-1}}, about \SI{500}{h} at nominal beam and target parameters. Solid black curve represents the reach for 1\% momentum resolution; dashed for 7\%; dotted for 15\%. Gray regions are existing leptonic exclusions, while colored regions show the range of parameters that could explain the respective anomalies. The blue shaded region shows the projected reach for the original Darklight detector design from earlier design studies.\cite{Kahn2012}}
\label{fig:reach}
\end{figure}

Figure~\ref{fig:reach} shows the resulting reach of the DarkLight experiment in the baseline and aggressive experimental designs described above, with a total integrated luminosity of \SI{1}{ab^{-1}}. The reach covers a majority of the $g_\mu-2$ preferred region under \SI{100}{MeV} and about half of the $^8$Be signal region\cite{Feng:2016ysn}. 

\section{Conclusion}
Using Geant4 simulations with realistic material assumptions, we demonstrated that a trigger system of helical fibers can be made sufficiently blind to elastic events to allow trigger operation for the DarkLight experiment. This design suppresses false triggering due to background events by two orders of magnitude with 30\% of signal efficiency at \SI{10}{MHz} readout rate at the designed luminosity. 

We further explored using helical-shaped fibers as possible lepton trackers. This concept was demonstrated using a simple track reconstruction method and produces a reach that covers the majority of the phase space of interest. We also estimated the resolution needed to obtain a given reach in the parameter space of $A'$. This design using scintillating fibers, which would be a cost efficient alternative to the original design of using GEM detectors, yields a slightly smaller but comparable reach\cite{Balewski2014}. These results motivate the further study of the sensitivity with more sophisticated tracking and the inclusion of pile-up and other considerations.

We note that this study demonstrated the feasibility of these two designs at the simulation level only. We did not take into account possible challenges in realizing these designs, including but not limited to the structural support for the fiber geometries, effects of the limited precision with which fibers can be positioned, light attenuation in the fibers, and a detailed treatment of other sources of noise. However, the results presented in this study give a solid numerical foundation for, and demonstrate an upper bound on, the performance of such a geometry. These results strongly encourage further efforts in prototype studies and extension to other similar electron proton scattering experiments.

\section*{Acknowledgements}
  This research is supported by the U.S. Department of Energy Office of Nuclear Physics under grant number DE-FG02-94ER40818.

\section*{References}
    \bibliography{main}

\end{document}